\documentclass[conference,letterpaper]{IEEEtran}

\usepackage{cite}
\usepackage{amsmath,amssymb,amsfonts,mathtools}
\usepackage{algorithm,algpseudocode}
\usepackage{graphicx}
\usepackage{textcomp}
\usepackage{xcolor}
\usepackage{bm}
\usepackage{float}
\usepackage[acronym,shortcuts]{glossaries}
\usepackage[english]{babel}
\usepackage{subcaption}

\usepackage[T1]{fontenc}

\newacronym{PDF}{PDF}{probability density function}
\newacronym{CR}{CR}{cognitive radio}
\newacronym{SotA}{SotA}{state of the art}
\newacronym{OFDM}{OFDM}{orthogonal frequency division multiplexing}
\newacronym{PAPR}{PAPR}{peak-to-average power ratio}
\newacronym{IFFT}{IFFT}{inverse fast Fourier transform}
\newacronym{DFT}{DFT}{discrete Fourier transform}
\newacronym{SQNR}{SQNR}{signal-to-quantization-noise ratio}
\newacronym{ADC}{ADC}{analog-to-digital converter}
\newacronym{DAC}{DAC}{digital-to-analog converter}
\newacronym{TR}{TR}{tone reservation}
\newacronym{RT}{RT}{reserved tone}
\newacronym{PRT}{PRT}{peak-reserved tone}
\newacronym{LP}{LP}{linear programming}
\newacronym{SLM}{SLM}{selective mapping}
\newacronym{DVB-T2}{DVB-T2}{DVB-second generation}
\newacronym{DVB-NGH}{DVB-NGH}{DVB for next generation handheld}
\newacronym{QCQP}{QCQP}{quadratically constrained quadratic program}
\newacronym{FP}{FP}{fractional programming}
\newacronym{CCDF}{CCDF}{complementary cumulative distribution function}
\newacronym{CDF}{CDF}{cumulative distribution function}
\newacronym{ISAC}{ISAC}{integrated sensing and communication}
\newacronym{FD}{FD}{frequency-domain}
\newacronym{TD}{TD}{time-domain}
\newacronym{QAM}{QAM}{quadrature amplitude modulation}
\newacronym{PSK}{PSK}{phase shift keying}
\newacronym{SNR}{SNR}{signal-to-noise ratio}
\newacronym{ACF}{ACF}{auto-correlation function}
\newacronym{Rx}{Rx}{receiver}
\newacronym{CP}{CP}{cyclic prefix}
\newacronym{RPE}{RPE}{radar parameter estimation}
\newacronym{ISI}{ISI}{inter-symbol interference}
\newacronym{MA}{MA}{multiple access}
\newacronym{OFDMA}{OFDMA}{orthogonal frequency division multiple access}
\newacronym{MMSE}{MMSE}{minimum mean square error}
\newacronym{QT}{QT}{quadratic transform}
\newacronym{KKT}{KKT}{Karush-Kuhn-Tucker}
\newacronym{MUI}{MUI}{multi-user interference}
\newacronym{OTAC}{OTAC}{over-the-air computing}
\newacronym{MSE}{MSE}{mean square error}
\newacronym{ATSC}{ATSC}{advanced television systems committee}
\newacronym{SINR}{SINR}{signal-to-interference-plus-noise ratio}

\def\BibTeX{{\rm B\kern-.05em{\sc i\kern-.025em b}\kern-.08em
T\kern-.1667em\lower.7ex\hbox{E}\kern-.125emX}}

\begin{document}
\glsresetall

\title{Joint PAPR Reduction and Over-the-Air Computing in Multi-User OFDMA Systems}

\author{%
  \IEEEauthorblockN{%
    Getuar Rexhepi\textsuperscript{*},
    Kuranage Roche Rayan Ranasinghe\textsuperscript{*},
    Giuseppe~Thadeu~Freitas~de~Abreu\textsuperscript{*},
    and David Gonz\'alez~G.\textsuperscript{\dag}}
  \IEEEauthorblockA{%
    \textsuperscript{*}\emph{School of Computer Science and Engineering}, \emph{Constructor University}, Bremen, Germany\\
    \textsuperscript{\dag}\emph{Wireless Communications Technologies Group}, \emph{AUMOVIO Germany GmbH}}
}

\maketitle
\begin{abstract}
We consider the \ac{PAPR} reduction challenge of \ac{OFDM} systems utilizing \ac{TR} under an over-the-air computing constraint, such that the signals placed in the reserved tones simultaneously enable distributed computation at the receiver.
To that end, the problem is cast as a fractional programming problem and solved via the iterative \ac{QT}, which guarantees convergence to a stationary point.
A key contribution is the decomposition of the reserved-tone vector into an orthogonal pair of components: one satisfying the sum-computation constraint and the other providing the remaining degrees of freedom for \ac{PAPR} reduction.
This decomposition reveals that the proposed method and the benchmark sacrifices the same number of degrees of freedom.
However, the proposed design distributes the computational payload across all reserved tones, yielding superior robustness against multi-user interference and quantization noise.
Analytical and simulation results confirm that the proposed method matches the benchmark in \ac{PAPR} reduction while outperforming it in computation accuracy, particularly in multi-user scenarios.
\end{abstract}

\glsresetall

\section{Introduction}
\label{sec:intro}

\Ac{OFDM} is one of the most widely used modulation techniques in modern wireless communication systems due to its robustness against multipath fading, its ability to suppress \ac{ISI}, and its seamless integration with \ac{MA} schemes \cite{Molisch2011}.
However, a well-known drawback of \ac{OFDM} is its high \ac{PAPR} \cite{ParedesG15}, whose significance grows with the number of subcarriers involved in the \ac{IFFT} during modulation.
A large \ac{PAPR} degrades the \ac{SQNR} of both the \ac{ADC} and the \ac{DAC}, and significantly reduces the efficiency of power amplifiers at the transmitter \cite{Udaigiriya2015}.
The goal of \ac{PAPR} reduction techniques is to lower the distance between the signal's peak and its average in the time domain without distorting its frequency spectrum.
Over the past two decades, many techniques have been developed, including coding, clipping, constellation shaping, \ac{SLM}, and \ac{TR} \cite{WenIWCMC2008, Tellado1999, JiangTB2008, RahmatallahCST2013, SelahattinTVT2019, TarakTB2021}.

Among these, the \ac{TR} technique \cite{Tellado1999}, which works via a subset of \ac{OFDM} subcarriers, the \acp{PRT}, reserved not for data but to carry peak-reducing signals, stands out for its simplicity and effectiveness.
Originally formulated as a convex \ac{LP} problem \cite{Tellado1999}, \ac{TR} has been adopted by standards including \ac{DVB-T2} \cite{DVB-T2Standard2012} and \ac{ATSC} 3.0 \cite{ATSC3_Standard2024}.
In conventional \ac{TR} methods \cite{LvMEC2013, LiTWC2020, BulusuITBC2018}, all \acp{PRT} are used solely for \ac{PAPR} reduction.

With the emergence of \ac{OTAC} as a promising technique for low-latency distributed computation in wireless networks \cite{GoldenbaumTCOM2013}, there is growing interest in integrating communication and computing functionalities within the same framework.
Since the reserved tones are not used for data transmission, they present an opportunity to embed computation results without affecting the data-carrying tones, potentially enabling a joint \ac{PAPR} reduction and over-the-air computing framework.
To the best of our knowledge, the possibility of embedding computation into the reserved tones of an \ac{OFDM} \ac{TR} framework has not been addressed in the open literature.

In this paper, we bridge this gap by proposing a joint \ac{PAPR} reduction and over-the-air computing framework for multi-user \ac{OFDMA} systems.
The proposed method exploits the reserved tones to simultaneously reduce the \ac{PAPR} and enable the receiver to compute a desired function of the users' local data.
Utilizing the \ac{QT} as a powerful tool for solving non-convex fractional programming problems \cite{Shen2018}, we reformulate the joint optimization problem into a tractable form that can be solved efficiently with guaranteed convergence to a stationary point.

A key insight is a structural decomposition of the reserved-tone vector that separates the computation degree of freedom from the \ac{PAPR}-reduction degrees of freedom, leading to a reduced optimization problem with the same number of free variables as the benchmark single-tone method, yet with superior performance due to the distributed computation across all reserved tones.
In addition, the proposed method completely eliminates \ac{MUI} at the receiver, allowing the computed result to be retrieved via a simple summation of the received reserved tones, without any additional combining step, thus reducing receiver complexity.
Simple algebraic manipulations show also that the method can perform multifunctional computing with the same reserved tones, e.g., computing different functions for different subsets of users, without any performance degradation.
Analytical and simulation results confirm that the proposed method achieves the same \ac{PAPR} reduction as the benchmark, while significantly outperforming it in computation accuracy.

\section{System Model}
\label{sec:system_model}

\subsection{OFDMA Signal Model}
Consider a multi-user \ac{OFDMA} system with $N$ subcarriers and $U$ users, where each
user is allocated a subset of subcarriers for data transmission.
Let $\bm{x}_u\in\mathbb{C}^{N\times 1}$ be the frequency-domain information vector of
user $u$, whose entries are complex symbols drawn from an arbitrary constellation
(e.g., $M$-QAM).
The \ac{TD} transmit signal of user $u$ is obtained via the \ac{IFFT}:
\begin{equation}
\label{eq:OFDM}
\mathbf{x}_u = \mathbf{F}_N^{\mathrm{H}} \bm{x}_u \in \mathbb{C}^{N\times 1},
\quad \forall\, u = 1, \ldots, U,
\end{equation}
where $\mathbf{F}_N \in\mathbb{C}^{N\times N}$ is the normalized \ac{DFT} matrix and
$(\cdot)^\mathrm{H}$ denotes the conjugate transpose.
Since all $U$ users transmit simultaneously, their signals superimpose over the air, so that the received signal is the sum
\begin{equation}
\mathbf{y} = \sum_{u=1}^{U} \mathbf{H}_u \mathbf{x}_u + \mathbf{n} \in \mathbb{C}^{N},
\end{equation}
where $\mathbf{H}_u\in\mathbb{C}^{N\times N}$ is the diagonal channel matrix of user $u$, with i.i.d.\ diagonal entries drawn from $\mathcal{CN}(0,1)$, and $\mathbf{n}\in\mathbb{C}^{N}$ has i.i.d.\ entries drawn from
$\mathcal{CN}(0,\sigma^2)$.
This superposition is precisely the mechanism exploited for over-the-air computing on the reserved tones (Section~\ref{sec:analysis}).

\begin{figure}
    \centering
    \includegraphics[width=\columnwidth]{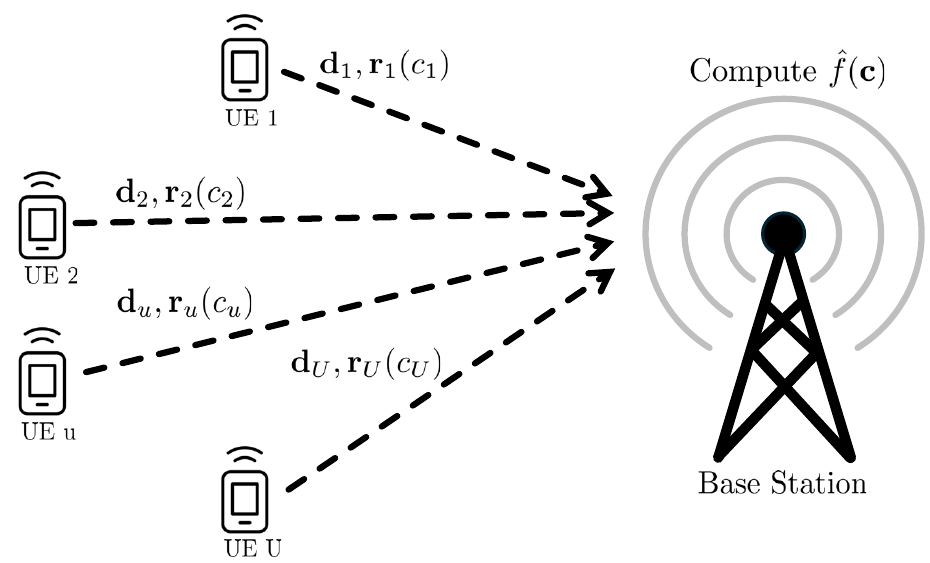}
    \caption{System model of the proposed multi-user OFDMA framework with joint PAPR reduction and over-the-air computing.}
    \label{fig:system_model}
\end{figure}

\subsection{PAPR Definition}
\Ac{PAPR} is a critical metric in \ac{OFDM} systems, quantifying the fluctuations in the signal's envelope and its impact on hardware performance.
The instantaneous \ac{PAPR} of a \ac{TD} signal $\mathbf{x}$ composed of $N$ discrete
samples is defined as 
\begin{equation}
\label{eq:PAPR}
\mathrm{PAPR}(\mathbf{x}) \triangleq
\frac{\|\mathbf{x}\|_{\infty}^2}{\frac{1}{N}\|\mathbf{x}\|_2^2},
\end{equation}
where $\|\cdot\|_{\infty}$ and $\|\cdot\|_2$ denote the $\ell_\infty$ and $\ell_2$
norms, respectively.

A high \ac{PAPR} indicates large signal peaks relative to the average power, leading to
inefficiencies in power amplifiers and increased distortion.
Effective \ac{PAPR} reduction techniques are essential to mitigate these issues, enabling more efficient and reliable communication, as well as supporting the integration of additional functionalities such as over-the-air computing.

\subsection{Tone Reservation for PAPR Reduction}
\label{sec:TR}

In the \ac{TR} framework \cite{Tellado1999}, $N_r$ subcarriers are reserved and used to
generate a peak-reducing signal added to the original transmit signal.
Decomposing the \ac{TD} signal into a data component $\mathbf{d} = \mathbf{F}^H\bm{x}_d$
and a reserved-tone component $\mathbf{F}_r^H \bm{r}$, the \ac{PAPR} minimization problem is formulated as
\begin{align}
  \min_{\bm{r} \in \mathbb{C}^{N_r}} \quad &
  \frac{\bigl\|\mathbf{d} + \mathbf{F}_r^H \bm{r}\bigr\|_{\infty}^2}
       {\frac{1}{N}\bigl\|\mathbf{d} + \mathbf{F}_r^H \bm{r}\bigr\|_{2}^2}
  \label{eq:TR}\\
  \text{s.t.}\quad & \|\bm{r}\|^2 \le P_{\max}, \nonumber
\end{align}
where $\bm{r}\in\mathbb{C}^{N_r}$ contains the symbols on the reserved subcarriers,
$P_{\max}$ is the maximum reserved-tone power, and $\mathbf{F}_r^H$ is the
sub-matrix of $\mathbf{F}^H$ corresponding to the reserved subcarriers.

Note that the denominator accounts for power variations inherent to QAM modulations; 
for PSK, the average power is constant, reducing the denominator to a scalar.
While numerous algorithms have been proposed to solve \eqref{eq:TR} 
\cite{Tellado1999, LvMEC2013, LiTWC2020, BulusuITBC2018}, they share a common 
limitation: optimizing only the numerator of the \ac{PAPR} expression, which is 
suboptimal for QAM constellations where the average power varies.
Despite using for other objectives, the solution to this problem was proposed by \cite{Rexhepi2025} to limit the number of active reserved tones, which will be useful for the proposed method in Section~\ref{sec:proposed}.

\subsection{Over-the-Air Computation}

\Ac{OTAC} exploits the superposition property of the wireless
multiple-access channel to allow multiple devices to transmit
simultaneously, so that the receiver recovers a {function} of the
aggregated data rather than the individual transmissions
\cite{GoldenbaumTCOM2013, RanasingheICNC2025,RanasingheArxiv2506}.
The class of functions naturally supported by this paradigm is that of
\emph{nomographic functions}, which admit
the general representation
\begin{equation}
    f\!\left(s_1,\dots,s_{N_r}\right)
    = \varphi\!\Bigl(\textstyle\sum_{i=1}^{N_r}\psi_i(s_i)\Bigr),
    \label{eq:nomographic}
\end{equation}
where $\psi_i:\mathbb{C}\to\mathbb{C}$ is a \emph{pre-processing}
(encoding) function applied locally at the $i$-th transmitter before
transmission, and $\varphi:\mathbb{C}\to\mathbb{C}$ is a
\emph{post-processing} (decoding) function applied at the receiver after
the over-the-air aggregation step.

This two-stage structure - local pre-processing followed by channel
superposition followed by global post-processing - is sufficiently
expressive to encompass a wide variety of practically relevant
aggregation tasks \cite{JesusCOMMST2014}, including the arithmetic mean, weighted sum, geometric mean, polynomial, and Euclidean norm, among others.

For non-linear targets, the pre- and post-processing pair $(\psi_i,
\varphi)$ must be designed jointly to ensure that the desired function
is recovered from the superimposed received signal.
As a prototypical example, the product function
$\prod_{i=1}^{N_r} r_i$ is rendered nomographic via the
log-domain linearization
\begin{equation}
    \prod_{i=1}^{N_r} r_i
    = \exp\!\Bigl(\bm{1}^H\log(\bm{r})\Bigr),
\end{equation}
where $\psi_i(r_i) = \log(r_i)$ and $\varphi(\cdot)=\exp(\cdot)$
serve as the pre- and post-processing maps, respectively.
More generally, as shown in \cite{GoldenbaumTCOM2013}, the framework in \eqref{eq:nomographic} is systematically generalizable
to any nomographic function by an appropriate choice of the pair
$(\psi_i, \varphi)$, and it scales gracefully to a massive number of
participating devices.

In the special and widely studied case of the \emph{linear sum}
function, one sets $\psi_i = \mathrm{id}$ (identity) and
$\varphi = \mathrm{id}$, so that \eqref{eq:nomographic} collapses to
\begin{equation}
    f(\bm{r}) = \sum_{i=1}^{N_r} r_i.   
\end{equation}
Under this choice, both the pre- and post-processing operations become
trivial, and the nomographic structure \eqref{eq:nomographic} reduces
exactly to the linear aggregation constraint $\bm{1}^H\bm{r} = c$,
$c\in\mathbb{C}$.
Consequently, following the convention adopted in
\cite{RanasingheICNC2025},
the explicit notation of $\psi_i$ and $\varphi$ can be omitted without
loss of generality when the target function is the linear sum, and the
\ac{OTAC} constraint is stated directly as $\bm{1}^H\bm{r} = c$.

\section{Proposed Method}
\label{sec:proposed}

\subsection{Benchmark: Single-Tone Embedding}
Before introducing the proposed method, we first describe a simple benchmark approach that embeds the computation result into a single reserved tone, while using the remaining reserved tones solely for \ac{PAPR} reduction.
To see this, we partition the reserved-tone vector as
\begin{equation}
  \bm{r} = [r_1, r_2, \ldots, r_{N_r-1},\, c]^T \in \mathbb{C}^{N_r},
\end{equation}
so that $c$ occupies one reserved tone and the remaining $N_r-1$ tones are free for
\ac{PAPR} reduction.
This sacrifices one degree of freedom and concentrates the computation payload on a
single subcarrier, making the computation result vulnerable to \ac{MUI} and quantization noise, especially in multi-user scenarios where the channel gains of the reserved tone may differ significantly across users.

\subsection{TR with Distributed Computing}
Instead of using a single tone, we propose to distribute the computation symbols over all
reserved tones via the following optimization problem
\begin{align}
  \min_{\bm{r} \in \mathbb{C}^{N_r}} \quad &
  \frac{\bigl\|\mathbf{d}+\mathbf{F}_r^H \bm{r}\bigr\|_{\infty}^2}
       {\frac{1}{N}\bigl\|\mathbf{d}+\mathbf{F}_r^H \bm{r}\bigr\|_{2}^2}
  \label{eq:prop}\\
  \text{s.t.}\quad &\|\bm{r}\|^2 \le P_{\max}, \nonumber\\
               &\bm{1}^H \mathbf{H}_{N_r} \bm{r} = c. \nonumber
\end{align}
where the constraint $\bm{1}^H \mathbf{H}_{N_r} \bm{r} = c$ ensures that the sum of the reserved tones equals the desired computation result $c$.

To expose the structure of this problem, we decompose $\bm{r}$ as
\begin{equation}
  \bm{r} = \bm{p}\, c + \bm{Q}\bm{v},
  \label{eq:decomp}
\end{equation}
where $\bm{p} = (\bm{h}_{N_r})^{*}/\|\bm{h}_{N_r}\|^2$, with $\bm{h}_{N_r} \triangleq \operatorname{diag}(\mathbf{H}_{N_r})$ the vector of reserved-tone channel gains, is the minimum-norm vector realizing the sum computation, $\bm{Q}\in\mathbb{C}^{N_r\times(N_r-1)}$ is a matrix whose
columns form an orthonormal basis for the null space of $\bm{1}^H\mathbf{H}_{N_r}$, and $\bm{v}\in\mathbb{C}^{N_r-1}$
collects the free optimization variables used for \ac{PAPR} reduction.
Since $\bm{1}^H\mathbf{H}_{N_r}\bm{p} = 1$ and $\bm{p}$ is orthogonal to the columns of $\bm{Q}$, the two components in \eqref{eq:decomp} are orthogonal.
This decomposition explicitly allocates one degree of freedom to the computation
constraint (via $c$) and $N_r-1$ degrees of freedom to \ac{PAPR} reduction (via
$\bm{v}$), identical to the benchmark.
Substituting \eqref{eq:decomp} into \eqref{eq:prop} yields the reduced problem:
\begin{align}
  \min_{\bm{v} \in \mathbb{C}^{N_r-1}} \quad &
  \frac{\bigl\|\mathbf{d}+\mathbf{F}_r^H \bm{r}(\bm{v})\bigr\|_{\infty}^2}
       {\frac{1}{N}\bigl\|\mathbf{d}+\mathbf{F}_r^H \bm{r}(\bm{v})\bigr\|_{2}^2}\\
  \text{s.t.}\quad &\|\bm{r}(\bm{v})\|^2 \le P_{\max}. \nonumber
\end{align}

\noindent\textbf{Feasibility condition.}
The computation value $c$ must satisfy
\begin{equation}
  |c|^2 \le P_{\max} \|\bm{h}_{N_r}\|^2,
\end{equation}
to ensure the power constraint is not violated.
Any desired computation result can be mapped into this feasible range via an
appropriate normalization strategy.

\subsection{Fractional Programming Reformulation}
\label{sec:FP}

The fractional objective in \eqref{eq:prop} is non-convex in general, as it involves 
a ratio of functions that are jointly non-convex in $\bm{v}$. 
Direct minimization is therefore intractable, and standard convex solvers cannot be applied without reformulation.

To overcome this, we leverage the \ac{QT} framework of \cite{Shen2018}, which provides 
a systematic way to handle fractional objectives by introducing an auxiliary variable 
that decouples the numerator and denominator. 
Specifically, the \ac{QT} reformulates the original problem as a sequence of tractable subproblems, each of which is convex and admits an efficient closed-form update for the auxiliary variable, guaranteeing 
monotonic improvement of the objective at each iteration.

Denoting $\mathbf{x}(\bm{v}) = \mathbf{d}+\mathbf{F}_r^H\bm{r}(\bm{v})$, we first 
flip the fraction to cast the minimization as a maximization problem. 
Applying the \ac{QT}, the original fractional program is equivalently reformulated as the following 
auxiliary problem at iteration $t$:
\begin{align}
\max_{\bm{v}^{(t)}} \quad &
2\,\Re\!\bigl\{(\bm{q}^{(t)})^H \mathbf{x}^{(t)}\bigr\}
- (\bm{q}^{(t)})^H\bm{q}^{(t)} \bigl\|\mathbf{x}^{(t)}\bigr\|_{\infty}^{2}
\label{eq:FP}\\
\text{s.t.} \quad & \|\bm{r}(\bm{v}^{(t)})\|^2 \le P_{\max}, \nonumber
\end{align}
where the auxiliary vector $\bm{q}^{(t)} \in \mathbb{C}^{N}$ is updated in closed form as
\begin{equation}
\bm{q}^{(t)} = \frac{\mathbf{x}^{(t)}}{\|\mathbf{x}^{(t)}\|_{\infty}^{2}}.
\end{equation}

Intuitively, $\bm{q}^{(t)}$ acts as a surrogate for the optimal ratio at each
iteration, updated by evaluating the fraction at the current iterate.
This decoupling is the key enabler of the \ac{QT} approach: for a fixed $\bm{q}^{(t)}$, the
subproblem \eqref{eq:FP} becomes convex and quadratic in $\bm{v}^{(t)}$, and can
therefore be solved efficiently using standard solvers (e.g., CVX).
The alternating update between $\bm{v}^{(t)}$ and $\bm{q}^{(t)}$ is guaranteed to monotonically
increase the objective, and the algorithm converges to a stationary point of the 
original problem \cite{Shen2018}.

The overall procedure is summarized in Algorithm~\ref{alg:PAPRFP}.

\begin{algorithm}[H]
\caption{PAPR Reduction with Distributed Computing}
\label{alg:PAPRFP}
\begin{algorithmic}[1]
\Require Data signal $\mathbf{d}$, sub-DFT matrix $\mathbf{F}_r$,
         power limit $P_{\max}$, computation value $c$
\State Initialize $\bm{v}^{(0)}$ randomly, $t \leftarrow 0$
\State Compute $\mathbf{x}^{(0)} = \mathbf{d}+\mathbf{F}_r^H\bm{r}(\bm{v}^{(0)})$,
       $\;\bm{q}^{(0)} = \frac{\mathbf{x}^{(0)}}{\|\mathbf{x}^{(0)}\|_\infty^2}$
\Repeat
  \State Solve \eqref{eq:FP} for $\bm{v}^{(t+1)}$
  \State Update $\mathbf{x}^{(t+1)} = \mathbf{d}+\mathbf{F}_r^H\bm{r}(\bm{v}^{(t+1)})$
  \State Update $\bm{q}^{(t+1)} = \frac{\mathbf{x}^{(t+1)}}{\|\mathbf{x}^{(t+1)}\|_\infty^2}$
  \State $t \leftarrow t+1$
\Until{Converged}
\Ensure Optimal $\bm{v}^*$ and reserved tones $\bm{r}^* = \bm{p}\,c + \bm{Q}\bm{v}^*$
\end{algorithmic}
\end{algorithm}

\section{Performance Analysis}
\label{sec:analysis}

\subsection{Proposed Method}

We now turn to the receiver side and show that the proposed embedding structure 
enables remarkably simple computation retrieval, with no residual \ac{MUI}.
Focusing on the received reserved tones, the received signal at the reserved 
subcarriers aggregates the contributions of all $U$ users and can be written as
\begin{equation}
\mathbf{y}_{\mathrm{res}}
  = \sum_{u=1}^U \mathbf{H}_{\mathrm{res},u}\bm{r}_u + \mathbf{n}_{\mathrm{res}},
\end{equation}
where $\mathbf{H}_{\mathrm{res},u}$ is the channel matrix seen by user $u$ over the 
reserved subcarriers, $\bm{r}_u$ is the corresponding reserved-tone vector, and 
$\mathbf{n}_{\mathrm{res}}$ is the additive noise contribution over the reserved 
subcarriers.

The receiver retrieves the computed result by applying a simple all-ones combining 
vector, i.e., summing all received reserved tones
\begin{equation}
\hat{f}(\bm{c})
  = \bm{1}^T\mathbf{y}_{\mathrm{res}}
  = \sum_{u=1}^U c_u + \hat{\mathbf{n}}_{\mathrm{res}},
\label{eq:prop_result}
\end{equation}
where $\hat{\mathbf{n}}_{\mathrm{res}}=\bm{1}^T\mathbf{n}_{\mathrm{res}}$ is the 
effective noise after combining. The key observation is that the \ac{MUI} term 
vanishes \emph{exactly} from \eqref{eq:prop_result}, i.e.,
\begin{equation}
    \bm{1}^T \sum_{u=1}^U \mathbf{H}_{\mathrm{res},u}\bm{Q}\bm{v}_u = \bm{0},
\end{equation}
which follows directly from the structural constraint $\bm{1}^T\mathbf{H}_{\mathrm{res},u}\bm{Q}=\bm{0}$
enforced by construction during the optimization. 
This is not a statistical cancellation or an approximation, it is an \emph{algebraic} identity that holds 
exactly for every channel realization, every user, and every noise instance.

Crucially, this zero-\ac{MUI} property is achieved without any dedicated 
interference cancellation step at the receiver, such as successive interference 
cancellation or minimum mean-square error combining. 
The computation result $\hat{f}(\bm{c}) = \sum_{u=1}^U c_u$ is retrieved via a trivial summation, 
significantly reducing receiver complexity compared to conventional \ac{OTAC} 
schemes that require more involved combining strategies. 
Furthermore, since the \ac{MUI} is eliminated structurally rather than statistically, the retrieval 
accuracy is limited only by the channel noise $\hat{\mathbf{n}}_{\mathrm{res}}$, 
and not by residual interference, making the proposed method particularly robust 
in dense multi-user scenarios.

\subsection{Benchmark Method}

The single-tone embedding approach concentrates the entire computation embedding 
onto a single dedicated reserved tone, subcarrier $N_r$, shared by all $U$ users. 
While this simplifies the transmitter design, it places the full burden of 
interference suppression on the receiver, which must employ an \ac{MMSE} combiner 
to extract the desired computation result from the superposition of all users' 
signals.

Specifically, the received signal at the dedicated subcarrier is:
\begin{equation}
    y_{N_r} = \sum_{u=1}^{U} H_{N_r,u} r_u + n_{N_r},
\end{equation}
where $H_{N_r,u}$ is the channel coefficient of user $u$ at subcarrier $N_r$,
$r_u$ is the corresponding reserved tone scalar, and $n_{N_r}$ is the additive 
noise. The \ac{MMSE} combiner is designed to maximize the \ac{SINR} at the 
output, and its weight is given by:
\begin{equation}
  w_{\mathrm{MMSE}} =
\frac{\sum_{u=1}^{U} H_{N_r,u}^*}
       {\sum_{u=1}^{U} |H_{N_r,u}|^2 + \sigma^2},
\end{equation}
where $\sigma^2$ denotes the noise variance. Applying $w_{\mathrm{MMSE}}$ to 
$y_{N_r}$, the resulting estimate expands as:
\begin{align}
\hat{f}(\bm{c}) &=
\underbrace{\frac{\sum_u|H_{N_r,u}|^2}{\sum_u|H_{N_r,u}|^2+\sigma^2}c_u}_{\text{desired signal}}
+\underbrace{\frac{\sum_{j\neq u}H_{N_r,u}^* H_{N_r,j}}{\sum_u|H_{N_r,u}|^2+\sigma^2}c_j}_{\text{residual MUI}}
+\text{noise}.\nonumber
\end{align}

The decomposition above reveals a fundamental limitation of the single-tone 
embedding approach. 
The desired signal term is attenuated by a factor strictly 
less than one due to the noise regularization in the denominator, and more 
critically, the \ac{MUI} term does \emph{not} vanish in general. 
Indeed, the residual \ac{MUI} is non-zero whenever the channel coefficients $H_{N_r,u}$ 
are not perfectly aligned in phase, which is the typical case in practice.

\vspace{-2ex}
\section{Simulation Results}
\label{sec:sims}

Unless otherwise stated, we consider $N=512$ subcarriers, $N_r=64$ reserved tones,
and different numbers of users, specified in the figure captions.
The maximum reserved-tone power is set to $P_{\max}=N_r$, which means that on average every reserved tone has power 1. 
Data symbols are drawn from a 16-QAM constellation, channel coefficients from
$\mathcal{CN}(0,1)$, and computation symbols from $\mathcal{CN}(0,\sigma_c^2)$.

\subsection{PAPR Performance}

\begin{figure}[t]
  \centering
  \includegraphics[width=\columnwidth]{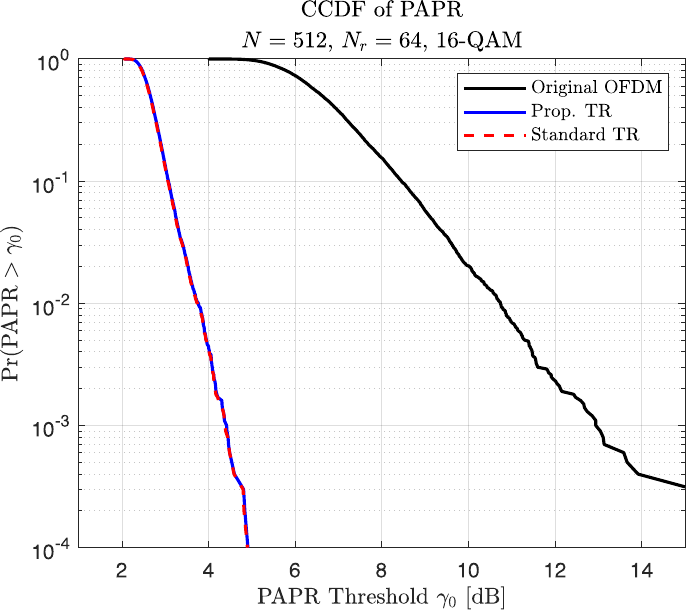}
  \caption{\ac{PAPR} reduction \ac{CCDF} comparison between the proposed and benchmark methods.}
  \label{fig:papr}
  \vspace{-3ex}
\end{figure}

Fig.~\ref{fig:papr} shows the \ac{CCDF} of the \ac{PAPR} for both methods.
As expected from the theoretical analysis in Section~\ref{sec:analysis}, both methods
sacrifice the same number of degrees of freedom for computation and thus achieve
equivalent \ac{PAPR} reduction performance.
This confirms that distributing the computation over all reserved tones does not
incur any \ac{PAPR} penalty compared to embedding it in a single tone.

\subsection{Computing Performance}
\vspace{-1ex}

\begin{figure}[t]
  \centering
  \includegraphics[width=\columnwidth]{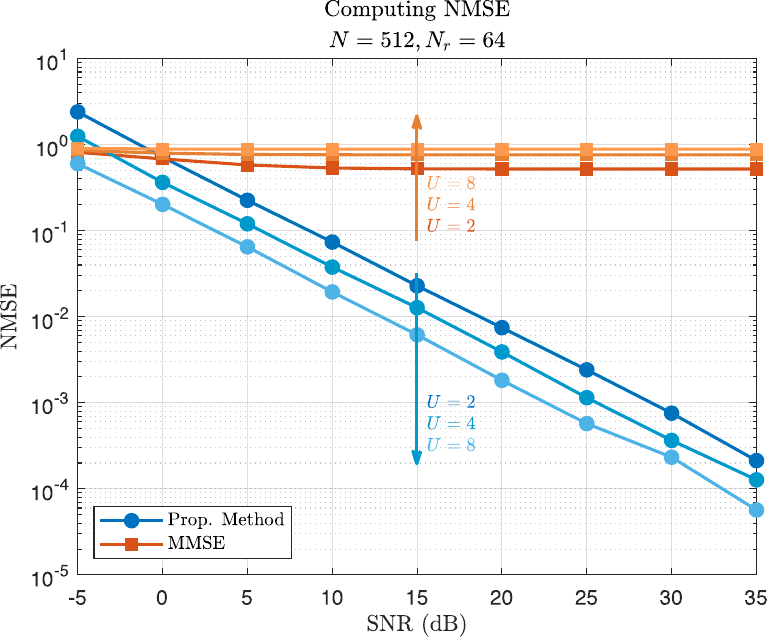}
  \caption{Computation \ac{MSE} versus \ac{SNR} for the proposed and benchmark methods, for $U = 2, 4, 8$ users.}
  \label{fig:computing}
  \vspace{-5ex}
\end{figure}

Fig.~\ref{fig:computing} compares the computation \ac{MSE} as a function of \ac{SNR}.
The proposed method significantly outperforms the benchmark across the entire \ac{SNR}
range, owing to its ability to completely eliminate \ac{MUI} through the orthogonal
decomposition of the reserved tones.
The benchmark method is limited by a residual \ac{MUI} floor that prevents it from
achieving arbitrarily low \ac{MSE} even at high \ac{SNR}.
%
\section{Conclusion}
\label{sec:conclusion}

We proposed a hardware-aware multi-user \ac{OFDMA} framework that jointly performs
\ac{PAPR} reduction and over-the-air computing via the \ac{TR} reserved tones.
A structural decomposition of the reserved-tone vector separates the computation
degree of freedom from the \ac{PAPR}-reduction degrees of freedom, yielding a
tractable fractional programming problem solved by a convergence-guaranteed
quadratic-transform algorithm.
The proposed method distributes the computation payload across all reserved tones,
completely eliminating multi-user interference at the receiver through a simple
summation, while achieving the same \ac{PAPR} reduction as the single-tone embedding
benchmark.
Future work will explore the extension of the proposed framework to more general nomographic functions, as well as its integration with advanced \ac{OFDMA} resource allocation strategies.

\bibliographystyle{IEEEtran}

\end{document}